\documentclass[showpacs,preprintnumbers,amsmath,amssymb]{revtex4}

\usepackage{graphicx}
\usepackage{dcolumn}
\usepackage{bm}
\usepackage{mathrsfs}
\usepackage{amsmath}

\begin{document}


\title{Perturbing microscopic black holes inspired by noncommutativity}

\author{D. Batic}
\email{davide.batic@ku.ac.ae}
\affiliation{%
Department of Mathematics,\\  Khalifa University of Science and Technology,\\ Main Campus, Abu Dhabi,\\ United Arab Emirates}
\author{N. G. Kelkar}
\email{nkelkar@uniandes.edu.co}
\affiliation{
Departamento de Fisica,\\ Universidad de los Andes, Cra.1E
No.18A-10, Bogota, Colombia
}

\author{M. Nowakowski}
\email{mnowakos@uniandes.edu.co}
\affiliation{
Departamento de Fisica,\\ Universidad de los Andes, Cra.1E
No.18A-10, Bogota, Colombia
}
\author{K. Redway}
\email{mnowakos@uniandes.edu.co}
\affiliation{
Departament of Physics,\\ University of the West Indies,\\ 
Mona Campus, Kingston\\
Jamaica
}%

\date{\today}

\begin{abstract}
We probe into the instabilities of microscopic quantum black holes. 
For this purpose, we study the quasinormal modes (QNMs) 
for a massless scalar perturbation of 
the noncommutative geometry inspired Schwarzschild black hole. 
By means of a sixth order Wentzel-Kramers-Brillouin (WKB) approximation we 
show that the widely used WKB method does not converge in the critical cases where 
instabilities show up at the third order. 
By employing the inverted potential method, we demonstrate that the instabilities are 
an artifact of the WKB method. Finally, we discuss the usefulness of the 
asymptotic iteration method to find the QNMs.
\end{abstract}

\pacs{04.70.-s,04.30.-w}
\maketitle

\section{Introduction}
The problem of understanding the mechanisms underlying the final evolution of a black hole, which, in turn, is closely related to the problem of the central singularity and the emergence of a possible black hole remnant, triggered several studies in the last two decades \cite{Adl}. Even though we do not have at the moment a final theory of quantum gravity able to offer a definite answer to the aforementioned problem, the existing candidate theories such as string theory, loop quantum gravity, and noncommutative geometry seem to share global features like the noncommutativity of the coordinates at a typical length $\sqrt{\theta}$ less than $10^{-18}m$ \cite{PO,Ma}, a new uncertainty principle including gravity effects \cite{Sa}, the avoidance of physical singularities \cite{Par}, and black hole remnants \cite{Gidd}. In particular, \cite{Piero1,Piero2,Piero3,Piero4} showed that noncommutativity can provide a cure for the central singularities afflicting the Schwarzschild and Reissner-Nordstr\"{o}m metrics. The corresponding noncommutative counterparts of these metrics are derived by incorporating noncommutativity only in the matter source while the Einstein action is kept unchanged. As a result of this approach, the central singularity is replaced by a regular region (deSitter core) represented by a self-gravitating, droplet of anisotropic fluid. The price to be paid is that the radial and tangential pressures are always negative, a fact which is difficult to explain. This difficulty was overcome in \cite{Davide1} where by means of a nonlocal equation of state a noncommutative mini black hole model was developed in which the pressure is positive in the interior of the droplet. Furthermore, \cite{us} gave the maximal singularity-free atlas for the noncommutative geometry inspired Schwarzschild metric \cite{Piero1}. This atlas describes an infinite lattice of asymptotically flat universes connected by black hole tunnels. The stability problem of the noncommutative Schwarzschild interior under massless scalar perturbations was attacked by \cite{DavideH,Mann} leading to conclusions opposite to ours. It is worth mentioning that \cite{An,Mo} claimed to have derived the noncommutative counterpart of the Kerr metric. Despite the educated guess leading to this result, the authors did not perform any consistency check regarding the question whether or not such a solution satisfies all Einstein field equations coupled to an anisotropic source. In this context, we mention the work of \cite{ABN} where it was shown that any noncommutative Kerr candidate metric, be it an educated guess or a particular solution, must satisfy a highly complicated system of 
partial differential equations (PDEs) represented by equations (12)-(19) therein. 
Moreover, Theorem~1 and 2 in \cite{ABN} give two independent proofs indicating that 
the metric found by \cite{An,Mo} can never satisfy all Einstein field equations and 
it may lead to contradictions. 

It is of general interest to study the stability of microscopic quantum black holes. 
Moreover, as it will become apparent in the text, there exists indeed a hint of 
possible instabilities when the third order WKB approximation is used to determine the 
QNMs. The main question we address here is if these instabilities persist by increasing 
the order of the WKB or by employing different algorithmic methods.
 
While there is a vast literature devoted to the study of QNMs for a commutative 
Schwarzschild spacetime \cite{QNMs}, the same cannot be said for the noncommutative 
Schwarzschild manifold for which only some partial results are available in the 
literature. For instance, \cite{Giri} analysed the asymptotic QNMs of a noncommutative 
geometry inspired Schwarzschild black hole under the assumption that the black hole 
mass $M$ is much larger than $M_e$, the mass of the extremal noncommutative 
Schwarzschild solution. At this point a comment is in order. From our Table~\ref{table:2} we can evince that the QNMs of a noncommutative Schwarzschild black hole already coincide with the corresponding ones of its classic counterpart when $M$ is of the same order of magnitude as $M_e$. Therefore, it is not surprising that \cite{Giri} reached the conclusion that the asymptotic QNMs of a noncommutative Schwarzschild black hole in the regime of large masses remain proportional to $\ln{3}$ as is the case for the classic Schwarzschild solution. Recently, \cite{Liang} studied the QNMs of massless scalar field perturbations 
in a noncommutative-geometry-inspired Schwarzschild black hole spacetime by means of the 
third-order WKB approximation. 
The author considered the cases with $\ell=1,2,3$ only and computed the QNMs for 
different values of the noncommutative parameter $\theta$ in the interval 
$0.01<\theta<0.2758$ using the third order WKB approximation.
Since the parameter $\sqrt{\theta}$ acts as a quantum of length, it is natural to 
assume $\sqrt{\theta}=\ell_P$ where $\ell_P=\sqrt{\hbar G/c^3}\approx 10^{-35}m$ is 
the Planck length. Then, the interval chosen in \cite{Liang} over which the 
noncommutative parameter varies corresponds to a choice of the black hole mass in the 
interval $3.6M_P<M<19M_P$ with $M_P$ the Planck mass. In this particular range 
studied by the author and within the third order WKB, results for the QNMs were 
found to be stable. The misleading instabilities occurring in the third order WKB, 
occur in the range $1.91M_P\leq M\leq 2.3897M_P$ as can be evinced from 
our Table~\ref{table:2}. Extending the WKB calculations up to the sixth order reveals 
that the method is not convergent exactly when the presumed instabilities occur. 
This forces us to turn to another reliable method to pin down the nature of these 
instabilities. We choose the inverted potential method by which we can assure that 
the instabilities are due to inefficiency of the WKB in such cases. This case teaches 
us an important lesson for the determination of QNMs: a single algorithm is not 
always appropriate to uncover all QNMs. We strengthen this conclusion by investigating 
the asymptotic iterative method (AIM) and show its advantage over other methods 
commonly used.

The paper is organised as follows. In Section II, we introduce a suitable rescaling 
of the noncommutative Schwarzschild manifold and derive the effective potential for a 
massless scalar field in the aforementioned geometry. Exploiting the short range property of this potential, we obtain a novel inequality linking the event horizon of a noncommutative Schwarzschild black hole with its mass parameter and the angular momentum quantum number of the scalar field. In Section III, we use a third and sixth order WKB approximation 
to compute the QNMs. Section IV discusses the inverted potential method followed by 
Section V devoted to the asymptotic iteration method.

\section{The noncommutative Schwarzschild black hole}
We consider a massless scalar field $\phi$ immersed in the noncommutative geometry inspired Schwarzschild background whose line element 
is \cite{Piero1} 
\begin{equation}\label{metric}
ds^2=f(r)dt^2-\frac{dr^2}{f(r)}-r^2\left(d\vartheta^2+\sin^2{\vartheta}d\varphi^2\right),\quad f(r)=1-\frac{2m(r)}{r} 
\end{equation}
with the mass function
\begin{equation}
m(r)=\frac{2M}{\sqrt{\pi}}\gamma\left(\frac{3}{2},\frac{r^2}{4\theta}\right), \quad 
\gamma\left(\frac{3}{2},\frac{r^2}{4\theta}\right)=\int_0^{r^2/4\theta}dt\sqrt{t} e^{-t},
\end{equation}
where $M$ is the total mass in the space-time manifold, $\theta$ is a parameter encoding noncommutativity and having the dimension of  a length squared, while $\gamma(\cdot,\cdot)$ is the incomplete lower gamma function. Taking into account that the incomplete and lower gamma functions are related by the formula \cite{abra}
\begin{equation}\label{conversion}
\gamma\left(\frac{3}{2},\frac{r^2}{4\theta}\right)+\Gamma\left(\frac{3}{2},\frac{r^2}{4\theta}\right)=\frac{\sqrt{\pi}}{2},
\end{equation}
the function $f(r)$ appearing in (\ref{metric}) can be written as the usual Schwarzschild counterpart plus a perturbation due to noncommutativity, i.e.
\begin{equation}
f(r)=1-\frac{2M}{r}+\frac{4M}{\sqrt{\pi}r}\Gamma\left(\frac{3}{2},\frac{r^2}{4\theta}\right).
\end{equation}
Observe that the classic Schwarzschild metric can be recovered in the regime $r^2/4\theta\gg 1$ because in that case $\Gamma(3/2,r^2/4\theta)\to 0$. Due to the spherical symmetry of the manifold described by (\ref{metric}), the Klein-Gordon equation
\begin{equation}
\square\phi=0,\quad \square=-\frac{1}{\sqrt{-g}}\partial_\mu(\sqrt{-g}g^{\mu\nu}\partial_\nu) 
\end{equation}
can be separated into spherical harmonics using the ansatz
\begin{equation}
\phi(t,r,\vartheta,\varphi)=\frac{\psi_{\omega\ell}(r)}{r}Y_{\ell m}(\vartheta,\varphi)e^{-i\omega t}. 
\end{equation}
with $\ell=0,1,2,\cdots$. According to \cite{Sam1}, the radial part $\psi_{\omega\ell}$ satisfies the following second-order, linear differential equation 
\begin{equation}\label{r1}
f(r)\frac{d}{dr}\left(f(r)\frac{d}{dr}\psi_{\omega\ell}(r)\right)+\left[\omega^2-U_{eff}(r)\right]\psi_{\omega\ell}(r)=0,\quad
U_{eff}(r)=f(r)\left[\frac{1}{r}\frac{df}{dr}+\frac{\ell(\ell+1)}{r^2}\right],
\end{equation}
where $U_{eff}$ is the effective potential. If we express the radial variable in units of the black hole mass and the black hole mass in units of the square root of noncommutative parameter, i.e
\begin{equation}
x=\frac{r}{M},\quad \mu=\frac{M}{\sqrt{\theta}},
\end{equation}
equation (\ref{r1}) can be cast into the form of an ordinary differential equation (ODE) 
characterized by two finite singularities at the inner and outer horizon and one singularity at infinity. More precisely, we find \begin{equation}\label{bigO}
f(x)\frac{d}{dx}\left(f(x)\frac{d}{dx}\psi_{\sigma\ell}(x)\right)+\left[\sigma^2-\mathcal{U}_{eff}(x)\right]\psi_{\sigma\ell}(x)=0,\quad
\mathcal{U}_{eff}(x)=f(x)\left[\frac{1}{x}\frac{df}{dx}+\frac{\ell(\ell+1)}{x^2}\right],
\end{equation}
where $\mathcal{U}_{eff}$ is the effective potential, $\sigma=M\omega$ is the dimensionless frequency defined as in \cite{IR}, and
\begin{equation}\label{f}
f(x)=1-\frac{2}{x}\mbox{erf}\left(\frac{1}{2}\mu x\right)+\frac{2\mu}{\sqrt{\pi}}e^{-\frac{1}{4}\mu^2 x^2},
\end{equation}
with $\mbox{erf}(\cdot)$ denoting the error function. The above expression for the function $f$ has been obtained by using the following relations in \cite{abra}, namely
\begin{equation}
\gamma\left(\frac{3}{2},z^2\right)=\frac{1}{2}\gamma\left(\frac{1}{2},z^2\right)-ze^{-z^2},\quad\gamma\left(\frac{1}{2},z^2\right)=\sqrt{\pi}\mbox{erf}(z).
\end{equation}
\begin{figure}\label{FIGON}
\includegraphics[scale=0.35]{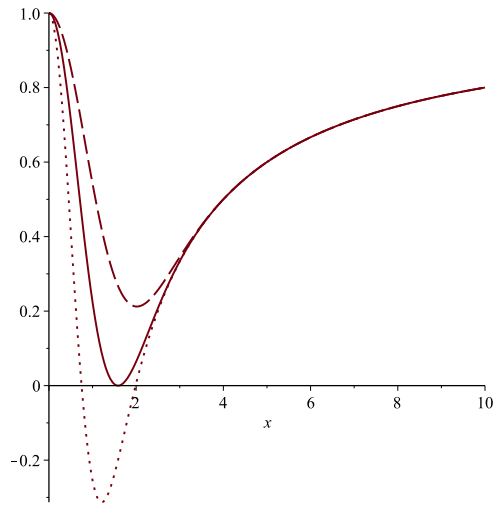}
\caption{\label{fig0}
Plot of the function $f(x)$ given by (\ref{f}). The solid line ($\mu_e= 1.904$) represents the scenario of an extreme black hole where the event and Cauchy horizons coalesce together at the common value $x_0= 1.587$. The dotted line ($\mu=2.5$) corresponds to the case of a non extreme black hole while the dashed line ($\mu=1.5$) describes a gravitational object characterized by a regular de Sitter core around $x=0$.}
\end{figure}
\begin{figure}
\includegraphics[scale=0.35]{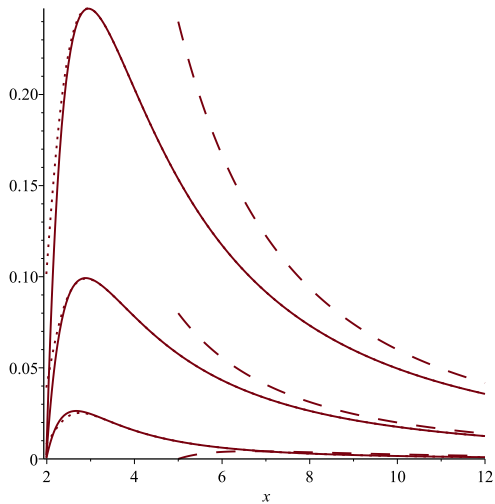}
\caption{\label{figX}
Plot of the effective potential $\mathcal{U}_{eff}$ in the case of an extreme (dotted line )and a non extreme noncommutative geometry inspired Schwarzschild black hole (solid line) for  different values of the angular momentum ($\ell=0$ bottom, $\ell=1$ middle, $\ell=2$ upper potential curve). The dashed line patterns represent the asymptotic approximations (\ref{exp1}) and (\ref{ap1a}). Approximation (\ref{exp1}) loses precision as $\ell$ increases.}
\end{figure}
\begin{figure}
\includegraphics[scale=0.35]{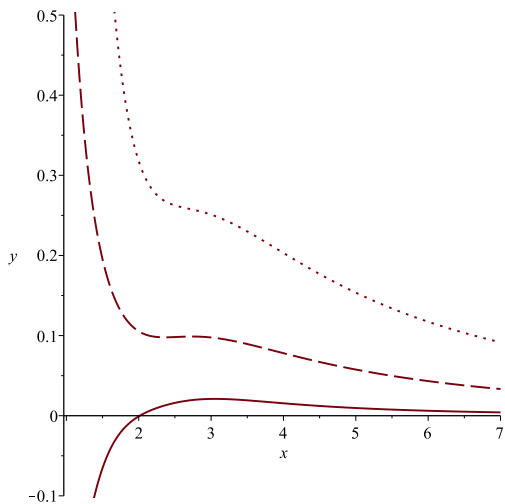}
\caption{\label{figX1}
Plot of the effective potential $\mathcal{U}_{eff}$ in the case of a naked de Sitter core with $\mu=1.5$ and different values of the angular momentum ($\ell=0$ line, $\ell=1$ dash, $\ell=2$ dot). The label on the y-axis refers to the effective potential.}
\end{figure}
\begin{figure}
\includegraphics[scale=0.35]{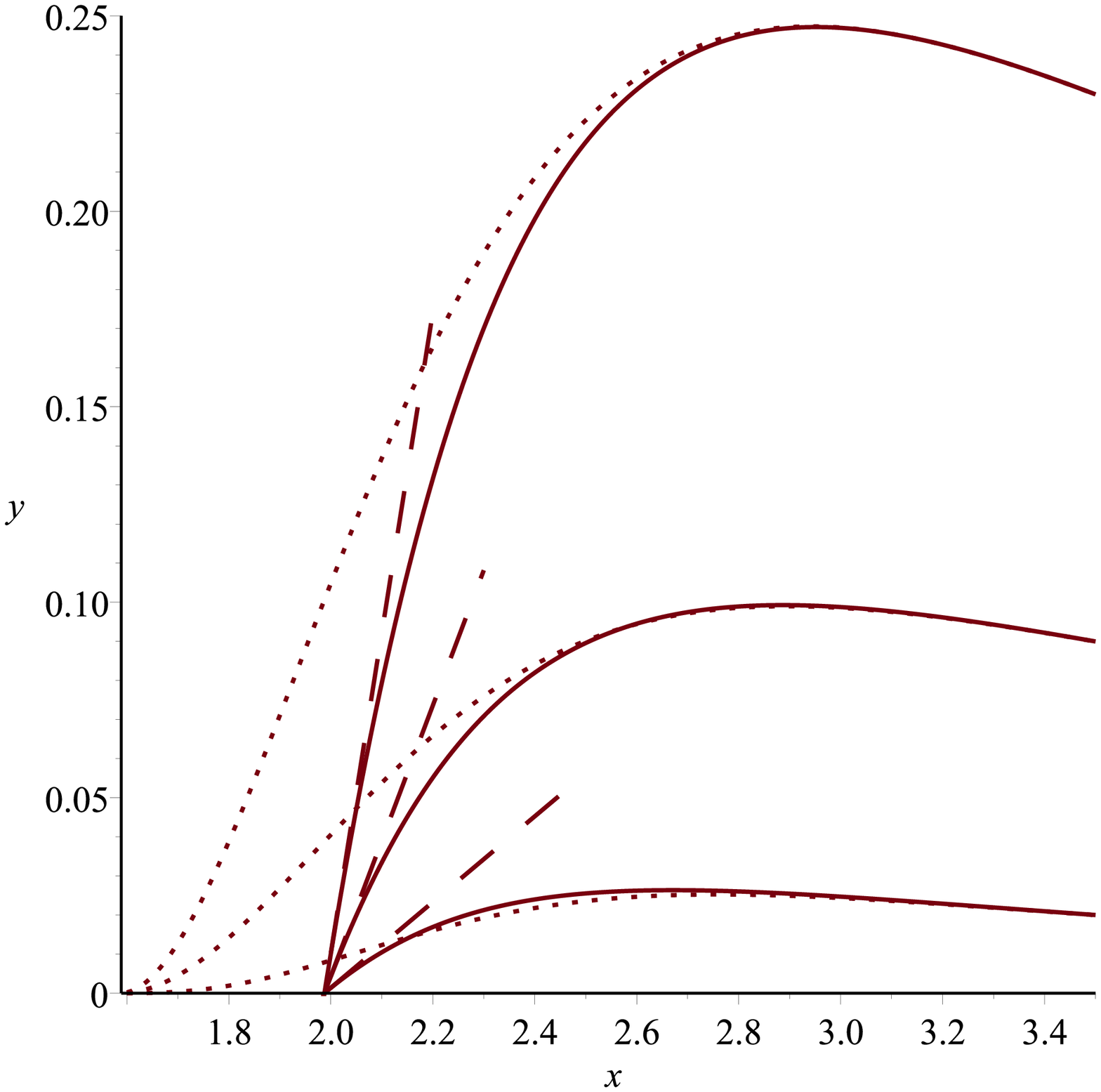}
\caption{\label{figB}
Comparison of the approximate form (\ref{exp2}) of the effective potential (dashed lines) close to the event horizon in the extreme case with the exact form of $U_{eff}$ (solid lines) given by (\ref{bigO}) for  different values of the angular momentum ($\ell=0$ bottom, $\ell=1$ middle, $\ell=2$ upper curves). The plot has been generated by expressing (\ref{exp2}) in terms of the variable $x$ with the help of (\ref{uppi}). The dotted lines represent the effective potential in the extreme case. Note the change in concavity at the event horizons for the effective potentials in the extreme and nonextreme cases. This behaviour signalizes that the approximation (\ref{exp2}) breaks down in the extreme case and it should be replaced by (\ref{ap2a}) as in Fig.~\ref{figA}. The label on the y-axis refers to the effective potential.}
\end{figure}
\begin{figure}
\includegraphics[scale=0.35]{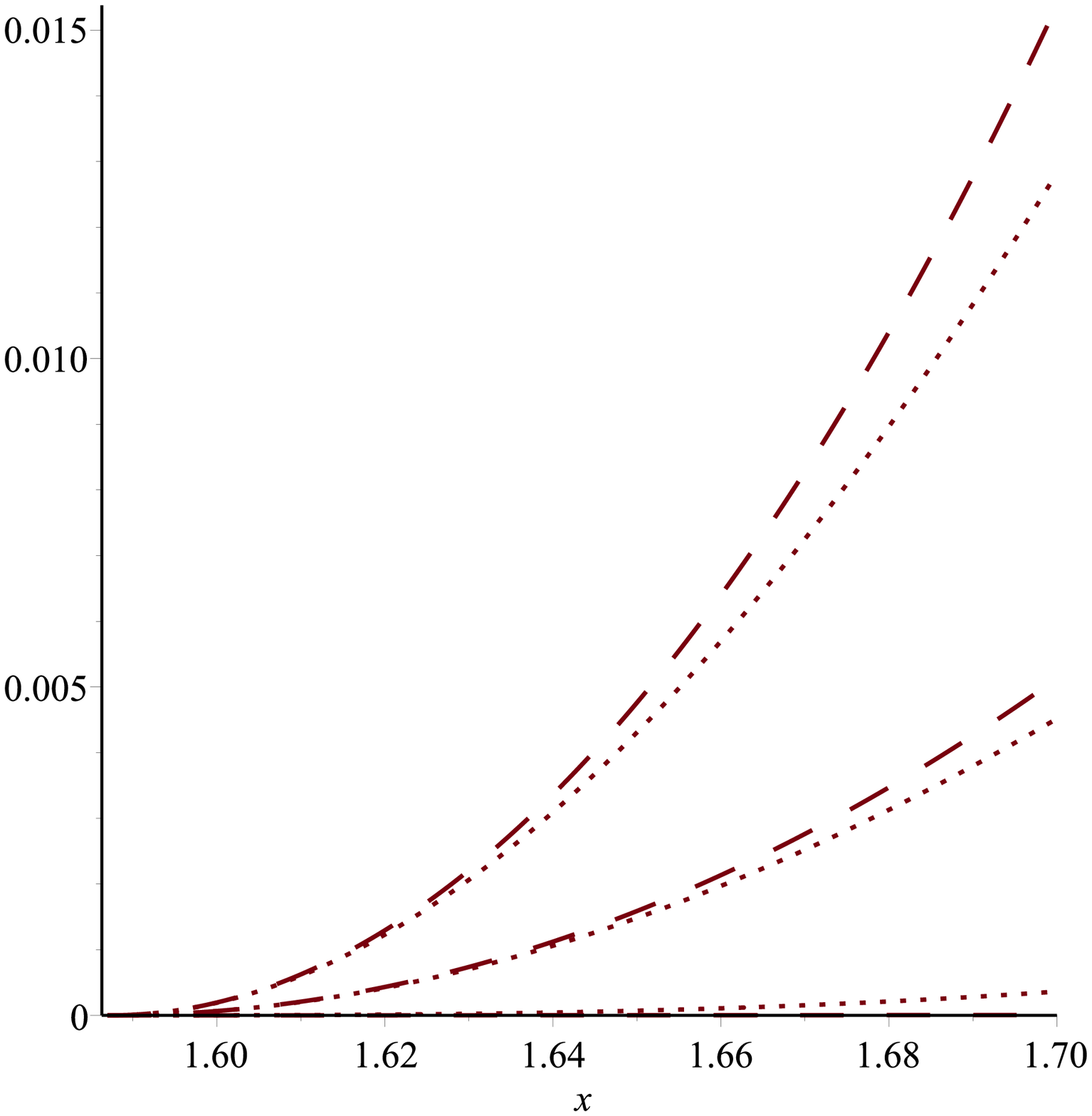}
\caption{\label{figA}
Comparison of the approximate form (\ref{ap2a}) of the effective potential (dashed lines) close to the event horizon in the extreme case with the exact form of $U_{eff}$ (dotted lines) given by (\ref{bigO}) for  different values of the angular momentum ($\ell=0$ bottom, $\ell=1$ middle, $\ell=2$ upper curves). The plot has been generated by expressing (\ref{ap2a}) in terms of the variable $x$ with the help of (\ref{uppi}).}
\end{figure}
It can be easily verified that an extreme noncommutative geometry inspired black hole will occur whenever the parameter $\mu$  has the critical value $\mu_e=1.904$. In this case, the Cauchy horizon $x_c$, and the event horizon $x_h$ coincide with $x_c=x_h=1.587$. For $\mu<\mu_{e}$, the line element (\ref{metric}) describes a naked gravitational object regular at $x=0$, while there are two distinct horizons if $\mu>\mu_{e}$. Furthermore, at short distances, i.e. $r\ll\sqrt{\theta}$, it turns out that the typical central singularity of a classic Schwarzschild black hole has been replaced by a regular deSitter core \cite{Piero1}. In the present work, we consider both the extreme and non extreme cases. The position of the event horizon $x_h$ is obtained by solving the transcendental equation $f(x_h)=0$, i.e.
\begin{equation}\label{EH}
x_h=2\mbox{erf}\left(\frac{1}{2}\mu x_h\right)-\frac{2\mu}{\sqrt{\pi}}x_h e^{-\frac{1}{4}\mu^2 x_h^2},
\end{equation}
Since the expression of the effective potential has an overall multiplicative term $f(x)$, we expect that $\mathcal{U}_{eff}$ vanishes at the event horizon. Furthermore, in the non extreme case the event horizon is a simple zero of the above equation, while it becomes a zero of order two in the extreme case \cite{us}. The differential equation (\ref{bigO}) can be further simplified by introducing the tortoise coordinate
\begin{equation}\label{def_tor}
\frac{dx_{*}}{dx}=\frac{1}{f(x)}.
\end{equation}
We observe that the function $x_{*}=x_{*}(x)$ is monotonically increasing on the interval $I=(x_h,\infty)$ because $f(x)$ is positive on $I$. Moreover, with the help of $7.1.23$ \cite{abra} it is not difficult to verify that asymptotically for $x\to+\infty$ the tortoise coordinate exhibits the behaviour  
\begin{equation}
x_{*}(x)=x+2\ln{x}+\mathcal{O}\left(\frac{1}{x}\right),\quad
x(x_{*})=x_{*}-2\ln{x_{*}}+\mathcal{O}\left(\frac{1}{x_{*}}\right).
\end{equation}
Since in the proximity of the event horizon, i.e. for $x\to x_h$, 
\begin{equation}
f(x)=f^{'}(x_h)(x-x_h)+\mathcal{O}(x-x_h)^2,
\end{equation}
the tortoise coordinate will behave there as
\begin{equation}\label{uppi}
x_{*}(x)=\frac{1}{\alpha}\ln{(x-x_h)}+\mathcal{O}(x-x_h),\quad\alpha=f^{'}(x_h).
\end{equation}
An inspection of (\ref{uppi}) shows that $x_{*}\to-\infty$ as $x\to x_h$ because the positive sign of $f^{'}(x_h)$ is ensured by the fact that $f(x)$ is increasing on the interval $(x_m,+\infty)$ with $x_m$ denoting the minimum of $f(x)$ (see dotted line in Fig.~\ref{FIGON}). Moreover, (\ref{uppi}) yields
\begin{equation}
x(x_{*})=x_h+e^{\alpha x_{*}}+\mathcal{O}\left(e^{2\alpha x_{*}}\right).
\end{equation}
In order to introduce the quasinormal modes boundary conditions, it is convenient to rewrite (\ref{bigO}) in terms of the tortoise coordinate. We get  
\begin{equation}\label{tor_eq}
\frac{d^2}{dx_{*}^2} \psi_{\sigma\ell}(x_{*})+\left[\sigma^2-\mathcal{U}_{eff}(x_{*})\right]\psi_{\sigma\ell}(x_{*})=0,\quad \mathcal{U}_{eff}(x_{*})=\frac{1}{x(x_{*})}\frac{df}{dx_{*}}+\ell(\ell+1)\frac{f(x(x_{*}))}{x^2(x_*)}.
\end{equation}
As in \cite{Sam1} solutions which are ingoing at the horizon (as $x_{*}\to-\infty$) satisfy
\[
\psi_{\sigma\ell}^{(in)}(x_{*})\sim
\left\{ \begin{array}{ll}
        e^{-i\sigma x_{*}} & \mbox{if $x_{*}\to-\infty$},\\
        A_{\sigma\ell}^{(in)}e^{-i\sigma x_{*}}+ A_{\sigma\ell}^{(out)}e^{i\sigma x_{*}}& \mbox{if $x_{*}\to+\infty$}.\end{array} 
\right.
\]
Quasi-normal modes are represented by the (complex) frequencies $\sigma_{\ell n}$, or equivalently $\omega_{\ell n}$ in virtue of the relation $\sigma=M\omega$, at which $A_{\sigma\ell}^{(in)}=0$. According to \cite{Nollert} the radial functions associated to the quasi-normal modes must diverge exponentially for $x_{*}\to\pm\infty$ and the wave function $\phi$ must exhibit an exponential decrease in the time variable. This requires that ${\rm{Im}}(\sigma)<0$. Note that if we would have taken $e^{i\omega t}$ in the separation ansatz for the Klein-Gordon equation, then one had to require ${\rm{Im}}(\sigma)>0$ together with the boundary conditions
\[
\lim_{x_{*}\to-\infty}\psi_{\sigma\ell}(x_{*})\sim e^{i\sigma x_{*}},\quad \lim_{x_{*}\to\infty}\psi_{\sigma\ell}(x_{*})\sim e^{-i\sigma x_{*}}.
\]
At this point a couple of remarks on the effective potential appearing in (\ref{tor_eq}) are in order. First of all, if we rewrite the lower incomplete Gamma function in terms of the upper incomplete Gamma function with the help of (\ref{conversion}), and then, we use $(2.24)$ in \cite{aslam} which provides an asymptotic expansion for the upper incomplete Gamma function, we discover that the asymptotic behaviour of the effective potential asymptotically away from the black hole is
\begin{equation}\label{exp1}
\mathcal{U}_{eff}(x_{*})=\frac{\ell(\ell+1)}{x^2_{*}}+\mathcal{O}\left(\frac{1}{x_{*}^3}\right).
\end{equation}
In the case $\ell=0$ the corresponding asymptotic expansion of the effective potential is given by
\begin{equation}\label{ap1a}
\mathcal{U}_{eff}(x_{*})=\frac{2\mu}{x^3_{*}}-\frac{8\mu}{\sqrt{\pi}}e^{-x^2_{*}}-\frac{4\mu}{x^4_{*}}+\mathcal{O}\left(\frac{e^{-x^2_{*}}}{x_{*}}\right).
\end{equation}
In Fig.~\ref{figX}, we compared (\ref{exp1}) and (\ref{ap1a}) with the exact shape of the effective potential. Moreover, when $x_{*}\to-\infty$, we find that at the event horizon of a nonextreme Schwarzschild black hole the effective potential can be approximated as follows
\begin{equation}\label{exp2}
\mathcal{U}_{eff}(x_{*})\sim\beta e^{\alpha x_{*}}+\mathcal{O}\left(e^{2\alpha x_{*}}\right),\quad
\beta=\frac{\alpha}{x_h}\left[\alpha+\frac{\ell(\ell+1)}{x_h}\right]
\end{equation}
with $\alpha$ given by (\ref{uppi}). As discussed  above,  $\alpha$ is always positive on the interval $(x_m,\infty)$, and therefore, $\beta>0$. The comparison of (\ref{exp2}) with the exact shape of the effective potential in a neighbourhood of the event horizon can be found in Fig.~\ref{figB}. In the extreme case, the approximation (\ref{exp2}) is not sharp enough due to a change in concavity of the effective potential near the event horizon as it can be evinced from Fig.~\ref{figB} and ~\ref{figA}, and therefore, it should be replaced by
\begin{equation}\label{ap2a}
\mathcal{U}_{eff}(x_{*})\sim\beta e^{\alpha x_{*}}+\gamma e^{2\alpha x_{*}}+\mathcal{O}\left(e^{3\alpha x_{*}}\right),\quad
\beta=\frac{\alpha}{x_h}\left[\alpha+\frac{\ell(\ell+1)}{x_h}\right],\quad\gamma=\frac{\beta f^{''}(x_h)}{2\alpha}-\frac{\alpha}{x_h}\left[\frac{\alpha}{x_h}-f^{''}(x_h)+\frac{2\ell(\ell+1)}{x^2_h}\right],
\end{equation}
in order to achieve a better fit with the behaviour of the effective potential in a neighbourhood of the event horizon. The asymptotic expansions (\ref{exp1})-(\ref{ap2a}) show that $\mathcal{U}_{eff}$ is a short-range potential. As such, $\mathcal{U}_{eff}$ must satisfy the condition \cite{chdet}
\begin{equation}\label{inizio}
\int^{+\infty}_{-\infty}\mathcal{U}_{eff}(x_{*})dx_{*}<\infty.
\end{equation}
Since the above integral can be computed analytically and the effective potential as a function of the tortoise coordinate is everywhere positive on the whole real line, we end up with the following inequality linking the event horizon of a  noncommutative geometry inspired Schwarzschild black hole with its mass and the angular momentum quantum number of the scalar field, namely
\begin{equation}\label{ungl}
x_h<\frac{2\ell(\ell+1)}{\mu^2}+\frac{2+\mu^2 x^2_h}{\mu^2 x_h}\mbox{erf}\left(\frac{1}{2}\mu x_h\right)-\frac{2}{\sqrt{\pi}\mu}e^{-\frac{1}{4}\mu^2 x^2_h}.
\end{equation}
The derivation of (\ref{ungl}) can be found in the Appendix~\ref{ineq}. It is interesting to observe that there is no counterpart of (\ref{ungl}) for a classic Schwarzschild black hole.

\section{A WKB approach}
We compute the quasinormal mode frequencies of a noncommutative geometry inspired 
Schwarzschild black hole in the presence of a massless scalar field by means of the 
WKB (Wentzel-Kramers-Brillouin) approximation carried to sixth order beyond the 
eikonal approximation \cite{konoplya}. On the way, we discover some shortcomings of 
the method, especially when applied to the noncommutative Schwarzschild case. 
In principle, the WKB approximation already at the third order is 
found to give reasonably accurate results 
\cite{schw,IR0,IR} in literature. However, in the noncommutative case under 
consideration, it can lead to misleading results. In order to bring forth the above 
point, we shall first present the WKB results at third order and then go over to 
show the limitations even at the sixth order of WKB.

To this purpose, we rewrite the radial equation (\ref{tor_eq}) describing the propagation of the field as
\begin{equation}\label{2.51}
\frac{d^2}{dx^2_{*}}\psi_{\sigma\ell}(x_{*})+Q(\sigma,x_{*})\psi_{\sigma\ell}(x_{*})=0,\quad Q(\sigma,x_{*})=\sigma^2-\mathcal{U}_{eff}(x_{*}),
\end{equation}
and make the ansatz
\begin{equation}\label{2.52}
\psi_{\sigma\ell}(x_{*})=A(\sigma,x_{*})e^{i\widehat{\phi}_{\sigma\ell}(x_{*})}
\end{equation}
with $A$ and $\widehat{\phi}$ functions to be determined. Substituting (\ref{2.52}) into (\ref{2.51}) gives
\begin{eqnarray}
2A^{'}(\sigma,x_{*})\widehat{\phi}_{\sigma\ell}^{'}(x_{*})+A(\sigma,x_{*})\widehat{\phi}_{\sigma\ell}^{''}(x_{*})&=&0,\label{2.53a}\\
A(\sigma,x_{*})\left[(\widehat{\phi}_{\sigma\ell}^{'}(x_{*}))^2-Q(\sigma,x_{*})\right]&=&A^{''}(\sigma,x_{*}),\label{2.53b}
\end{eqnarray}
where a prime denotes differentiation with respect to $x_{*}$. Suppose that $A^{''}\sim 0$. Recalling that in the case $|x_{*}|\to\infty$ the function $Q(\sigma,x_{*})$ tends to a constant, the solution of (\ref{2.51}) is
\begin{equation}\label{2.54}
\psi_{\sigma\ell}(x_{*})\sim e^{\pm i\sigma x_{*}},\quad\Re{(\sigma)}>0.
\end{equation}
It is clear that this will be the case when $Q(\sigma,x_{*})$ and $V_{eff}(x_{*})$ vary slowly with respect to the wave length $\lambda=2\pi/\sigma$. Under this assumption, it is reasonable to expect that $\psi_{\sigma\ell}(x_{*})$ is approximately given by (\ref{2.54}) even when $A(\sigma,x_{*})$ changes gradually, i.e. its variation is appreciable only on intervals of total length of the order of some wavelengths. Solving (\ref{2.53a}) and (\ref{2.53b}) under the aforementioned  assumption gives
\begin{equation}\label{2.55}
\psi_{\sigma\ell}(x_{*})\sim\frac{1}{\sqrt[4]{Q(\sigma,x_{*})}}e^{\pm i\int\sqrt{Q(\sigma,x_{*})}~dx_{*}}.
\end{equation}
Observe that the expression above will diverge when $Q(\sigma,x_{*})\sim 0$. This aspect turns out to be important  in the determination of the QNMs. Since $Q(\sigma,x_{*})$ tends to a constant as $|x_{*}|\to\infty$, the behaviour of the radial wave function will be captured by (\ref{2.54}). Furthermore, for a wave coming from space-like infinity and hitting the potential barrier over some interval of finite length only a fraction of the wave will be transmitted towards the event horizon and the transmitted wave will be reduced with respect to the incoming wave by a certain factor. In the case $\sigma\gg V_{eff}(x_{*})$ we expect instead full transmission. If $x_{*,m}$ denotes the maximum of $-Q(\sigma,x_{*})$, the first scenario described above will apply for $Q(\sigma,x_{*,m})<0$ while the second one for $Q(\sigma,x_{*,m})\sim 0$ \cite{schw}. There are some boundary conditions to be considered. First of all, we need to verify that for $x_{*}\to-\infty$ there are only incoming waves towards the event horizon. Furthermore, there must be no waves coming from $x_{*}\to+\infty$ For this reason, the interaction of the field with the potential peak must generate a reflected and a transmitted wave, both having the same order of magnitude. In light of the above discussion, we expect that the waves will exhibit characteristic frequencies such that $Q(\sigma,x_{*,m})\sim 0$. This implies that (\ref{2.55}) is not valid in some neighbourhood $(x_{*,1},x_{*,2})$ around $x_{*,m}$ and hence, it should be replaced by
\begin{equation}\label{2.56-57}
\psi_{\sigma\ell}(x_{*})\sim\frac{1}{\sqrt[4]{Q(\sigma,x_{*})}}\left\{
\begin{array}{cc}
\mbox{exp}\left(\pm i\int_{x_{*,2}}^{x_{*}}\sqrt{Q(\sigma,z)}~dz\right),&x_{*}>x_{*,2}\\
\mbox{exp}\left(\pm i\int_{x_{*}}^{x_{*,1}}\sqrt{Q(\sigma,z)}~dz\right),&x_{*}<x_{*,1}.
\end{array}
\right.
\end{equation}
At this point we still need to match (\ref{2.56-57}) with the solution on the interval $(x_{*,1},x_{*,2})$. The problem of matching simultaneously the two exterior WKB solutions across both turning points can be approached as in \cite{IR0,IR}. To this purpose, we consider a Taylor expansion of $Q(\sigma,x_{*})$ at the sixth order in a neighbourhood of the maximum $x_{*,m}$ in the inner region. This procedure allows to derive an asymptotic approximation to the interior solution which in turn can be used to connect the two WKB solutions in (\ref{2.56-57}). As a result, two connection formulas emerge that relate the amplitude of the incoming and outgoing wave on either side of the potential wall. The boundary condition of a purely outgoing wave leads to the following semi-analytical formula for the QNMs \cite{IR0,IR}
\begin{equation}\label{QNMS_formula}
\sigma^2=\mathcal{U}_{eff}(x_{*,m})+\sqrt{-2\mathcal{U}^{''}_{eff}(x_{*,m})}\Lambda(n)-i\left(n+\frac{1}{2}\right)\sqrt{-2\mathcal{U}^{''}_{eff}(x_{*,m})}\left[1-\Sigma(n)\right],\quad n=0,1,\cdots
\end{equation}
with
\begin{eqnarray*}
\Lambda(n)&=&\frac{1}{\sqrt{-2\mathcal{U}^{''}_{eff}(x_{*,m})}}\left[\frac{\mathcal{U}_{eff}^{(iv)}(x_{*,m})}{8\mathcal{U}^{''}_{eff}(x_{*,m})}\left(\frac{1}{4}+\gamma^2\right)-\frac{1}{288}\left(\frac{\mathcal{U}_{eff}^{'''}(x_{*,m})}{\mathcal{U}^{''}_{eff}(x_{*,m})}\right)^2(7+60\gamma^2)\right],\quad\gamma=n+\frac{1}{2},\\
\Sigma(n)&=&\frac{1}{2\mathcal{U}^{''}_{eff}(x_{*,m})}\left[\frac{5}{6912}\left(\frac{\mathcal{U}_{eff}^{'''}(x_{*,m})}{\mathcal{U}^{''}_{eff}(x_{*,m})}\right)^4(77+188\gamma^2)-\frac{{\mathcal{U}^{'''}}^2_{eff}(x_{*,m})\mathcal{U}_{eff}^{(iv)}(x_{*,m})}{384{\mathcal{U}^{''}}^3_{eff}(x_{*,m})}(51+100\gamma^2)+\right.\\
&&\left.\frac{1}{2304}\left(\frac{\mathcal{U}_{eff}^{(iv)}(x_{*,m})}{\mathcal{U}^{''}_{eff}(x_{*,m})}\right)^2(67+68\gamma^2)+\frac{\mathcal{U}_{eff}^{'''}(x_{*,m})\mathcal{U}_{eff}^{(v)}(x_{*,m})}{288{\mathcal{U}^{''}}^2_{eff}(x_{*,m})}(19+28\gamma^2)-\frac{\mathcal{U}_{eff}^{(vi)}(x_{*,m})}{288\mathcal{U}^{''}_{eff}(x_{*,m})}(5+4\gamma^2)\right],
\end{eqnarray*}
where ${}^{'}=d/dx_{*}$. Since the formulae for the higher order derivatives of the effective potential can be easily evaluated using Maple, they will not be presented here.
\begin{table}[ht]
\centering
 \begin{tabular}{||c| c| c| c| c| c| c|c||} 
 \hline
 $\ell$ & $n$ & $\sigma_{L}$ & $\sigma_{WKB}$ & $\sigma_e$ & $\sigma$, $\mu=1.95$ & $\sigma$, $\mu=2.25$ &$\sigma$, $\mu=100$\\ [0.5ex] 
 \hline\hline
 $0$ & $0$ & $0.1105-0.1049i$ & $0.1046-0.1152i$ & $0.0395+0.1367i$ & $0.0397+0.1330i$ & $0.0222-0.0944i$& $0.1046-0.1152i$\\ 
     & $1$ & $0.0861-0.3481i$ & $0.0892-0.3550i$ & $0.1524+0.4582i$ & $0.1555+0.4548i$ & $0.0880+0.3947i$& $0.0892-0.3550i$\\
 $1$ & $0$ & $0.2929-0.0977i$ & $0.2911-0.0980i$ & $0.2699-0.0744i$ & $0.2727-0.0818i$ & $0.2882-0.0975i$& $0.2911-0.0980i$\\
     & $1$ & $0.2645-0.3063i$ & $0.2622-0.3704i$ & $0.1061-0.2684i$ & $0.1256-0.2689i$ & $0.2449-0.3059i$& $0.2622-0.3704i$\\
     & $2$ & $0.2295-0.5401i$ & $0.2235-0.5268i$ & $0.1023+0.5721i$ & $0.0779+0.5589i$ & $0.1771-0.5284i$& $0.2235-0.5268i$\\ 
     & $3$ & $0.2033-0.7883i$ & $0.1737-0.7486i$ & $0.3374+0.9027i$ & $0.3072+0.8805i$ & $0.0862-0.7594i$& $0.1737-0.7486i$\\
 $2$ & $0$ & $0.4836-0.0968i$ & $0.4832-0.0968i$ & $0.4756-0.0870i$ & $0.4767-0.0889i$ & $0.4819-0.0964i$& $0.4832-0.0968i$\\
     & $1$ & $0.4639-0.2956i$ & $0.4632-0.2958i$ & $0.4136-0.2612i$ & $0.4209-0.2669i$ & $0.4561-0.2944i$& $0.4632-0.2958i$\\
     & $2$ & $0.4305-0.5086i$ & $0.4317-0.5034i$ & $0.2782-0.4540i$ & $0.2999-0.4590i$ & $0.4122-0.5015i$& $0.4317-0.5034i$\\
     & $3$ & $0.3939-0.7381i$ & $0.3926-0.7159i$ & $0.0918-0.6961i$ & $0.1285-0.6908i$ & $0.3544-0.7154i$& $0.3926-0.7159i$
     \\[1ex] 
 \hline
 \end{tabular}
 \caption{Normal modes for scalar perturbations of the noncommutative geometry inspired Schwarzschild metric. The third column represents the numerical values found by 
\cite{Leaver} in the case of a classic Schwarzschild black hole while the fourth columns 
reports the corresponding WKB results found by \cite{IR}. 
Here, $\sigma=M\omega$ is the dimensionless frequency emerging after having expressed the radial coordinate in units of the black hole mass $M$, and $\sigma_e$ denotes the QNM in the case of an extreme non-commutative geometry inspired Schwarzschild black hole.}
\label{table:0}
\end{table}
Table~\ref{table:0} shows that the extreme noncommutative geometry inspired Schwarzschild black hole is unstable, that is a positive imaginary part of the QNM leads to exponential growth of a scalar perturbation. Using different boundary conditions in the context of black plus mirror, \cite{Cardoso} also finds similar instability behaviour for small Kerr-AdS black holes. Furthermore, the non extreme case also exhibits instabilities when the mass parameter $\mu$ slightly exceed the corresponding mass parameter of an extreme Schwarzschild black hole. 
Table~\ref{table:2} also shows that for fixed $\ell$ but increasing $n$ the corresponding QNM of the noncommutative geometry inspired Schwarzschild black hole approaches its classical counterpart 
already for small values of the mass parameter. These results at third order WKB 
suggest that the effects due to 
noncommutative geometry are relevant for microscopic black holes but they can be 
neglected when dealing with black holes of astrophysical interest.
\begin{table}[ht]
\centering
 \begin{tabular}{||c| c| c| c| c| c| c| c| c||} 
 \hline
 $\mu$ & $\sigma,~\ell=n=0$ & $\mu$ & $\sigma,~\ell=0,~n=1$  & $\mu$ & $\sigma,~\ell=1,~n=2$ & $\mu$ & $\sigma,~\ell=1,~n=3$\\ [0.5ex] 
 \hline\hline
$1.9100$ & $0.0396+0.1363i$ & $1.9100$ & $0.1529+0.4579i$ & $1.9100$ & $0.0997+0.5705i$ & $1.9100$ & $0.3343+0.9001i$\\
$1.9500$ & $0.0396+0.1330i$ & $1.9500$ & $0.1555+0.4548i$ & $1.9500$ & $0.0779+0.5589i$ & $1.9500$ & $0.3072+0.8805i$\\
$2.0000$ & $0.0380+0.1279i$ & $2.3800$ & $0.0063+0.3670i$ & $2.0465$ & $0.0003+0.5334i$ & $2.1000$ & $0.1261+0.7960i$\\
$2.1958$ & $0.0000+0.0989i$ & $2.3890$ & $0.0003+0.3658i$ & $2.0467$ & $0.0000+0.5333i$ & $2.1800$ & $0.0052+0.7677i$\\
$2.1959$ & $0.0000-0.0989i$ & $2.3895$ & $0.0001+0.3658i$ & $2.0468$ & $0.0000-0.5333i$ & $2.1835$ & $0.0002+0.7669i$\\
$2.2000$ & $0.0016-0.0983i$ & $2.3897$ & $0.0000+0.3658i$ & $2.0500$ & $0.0030-0.5327i$ & $2.1836$ & $0.0000+0.7669i$\\ 
$2.6000$ & $0.1057-0.1189i$ & $2.3898$ & $0.0000-0.3657i$ & $2.1000$ & $0.0517-0.5252i$ & $2.1837$ & $0.0001-0.7668i$\\
$2.7500$ & $0.1097-0.1201i$ & $2.3900$ & $0.0001-0.3657i$ & $2.2000$ & $0.1422-0.5247i$ & $2.2500$ & $0.0862-0.7594i$\\
$2.8000$ & $0.1095-0.1196i$ & $2.3950$ & $0.0034-0.3651i$ & $2.4500$ & $0.2398-0.5376i$ & $2.3500$ & $0.1706-0.7605i$\\
$2.9000$ & $0.1083-0.1183i$ & $2.4000$ & $0.0066-0.3646i$ & $2.5500$ & $0.2417-0.5358i$ & $2.6500$ & $0.2001-0.7552i$\\
$3.1000$ & $0.1059-0.1161i$ & $2.4500$ & $0.0363-0.3611i$ & $2.6000$ & $0.2398-0.5343i$ & $2.8000$ & $0.1851-0.7510i$\\
$3.1500$ & $0.1051-0.1158i$ & $2.5000$ & $0.0611-0.3601i$ & $2.9500$ & $0.2252-0.5274i$ & $2.9000$ & $0.1789-0.7496i$\\
$3.2000$ & $0.1053-0.1156i$ & $3.0000$ & $0.0972-0.4566i$ & $3.1000$ & $0.2238-0.5269i$ & $3.0000$ & $0.1756-0.7489i$\\
$3.3000$ & $0.1049-0.1153i$ & $3.4500$ & $0.0893-0.3549i$ & $3.2000$ & $0.2236-0.5268i$ & $3.1000$ & $0.1742-0.7487i$\\
$4.0000$ & $0.1046-0.1152i$ & $4.0000$ & $0.0892-0.3550i$ & $3.2500$ & $0.2235-0.5268i$ & $3.2000$ & $0.1737-0.7486i$
     \\[1ex] 
 \hline
 \end{tabular}
 \caption{Transition from unstable to stable modes in the cases $\ell=0$, $n=0,1$, and $\ell=1$, $n=2,3$, using the WKB approximation at the third order.}
\label{table:2}
\end{table}

Extending the above calculations up to the sixth order, as mentioned above, we notice 
that the instabilities are not a real effect but rather an artifact or limitation 
of the WKB method. To demonstrate this limitation, in Table \ref{table3}, we 
show the QNMs for two values of $\mu$ which represent the extreme and non-extreme cases. 
We observe that the results do not converge when $l \le n$ and in fact the 
convergence gets better when $l \gg n$. Though the table displays the results up to 
the sixth order, for the cases where $l \le n$, there is no convergence found even 
up to the thirteenth order of WKB. To conclude this section, we note that though the 
WKB approximation has in general been found to lead to precise results when extended 
up to the 13th order, one should be cautious in the interpretation of new effects 
such as instabilities while using this method.

\begin{table}[ht]
\centering
 \begin{tabular}{||c| c| c| c| c| c| c| c||} 
 \hline
Order & 6 & 5 & 4  & 3 & 2 & Eikonal\\ [0.5ex] 
 \hline\hline
 $\mu = 1.91$\\ \hline
$l,n$\\ \hline
0, 0& $1.61847 + 0.06772 i$ & $0.36561 + 0.29978 i$ & $0.21084 + 0.02561 i$ & $0.03962 + 0.13627 i$ & $0.11332 - 0.17275 i$ & $0.18958 - 0.10326 i$ \\
1, 0 &$0.21155 - 0.30434 i$&$ 0.36645 - 0.17569 i$& $0.32747 - 
 0.06178 i$&$ 0.27024 - 0.07487 i$&$ 0.28245 - 
 0.11116 i$&$ 0.32879 - 0.09549 i$\\ 
2, 0& $0.45257 - 0.11220 i$& $0.48987 - 0.10365 i$&$0.48633 - 
 0.08538 i$&$0.47574 - 0.08728 i$&$ 0.47800 - 
 0.09886 i$& $0.50595 - 0.09340 i$\\
2, 1 & $0.38600 - 0.47357 i$&$0.54088 - 0.33797 i$&$0.47923 - 
 0.22658 i$&$0.41456 - 0.26193 i$& $0.44999 - 0.31504 i$& $0.55833 - 
 0.25391 i$\\
3, 0 &$0.66888 - 0.09442 i$& $0.67504 - 0.09356 i$& $0.67447 - 
 0.08938 i$& $0.67099 - 0.08985 i$& $0.67178 - 
 0.09552 i$& $0.69172 - 0.09277i$\\
3, 1& $0.62869 - 0.30765 i$& $0.66875 - 0.28923 i$&  $0.65638 - 
 0.25934 i$& $0.63424 - 0.26840 i$& $0.64709 - 0.29750 i$& $0.73394 - 
 0.26230 i$\\
3, 2 & $0.59901 - 0.57927 i$& $0.69462 - 0.49954 i$& $0.62507 - 
 0.39719 i$ &$0.55413 - 0.44803 i$& $0.61513 - 0.52160 i$&  $0.79536 - 
 0.40340 i$\\
\hline
$\mu = 2.7$\\ \hline
 0, 0&$0.84966 + 0.06067 i$& $0.20661 + 0.24951 i$& $0.08130 - 
 0.16179 i$& $0.10936 - 0.12028 i$& $0.13227 - 0.14143 i$&$ 0.18991 - 
 0.09851 i$ 
\\
1, 0 & $0.30052 - 0.08903 i$& $0.28808 - 0.09288 i$& $0.29024 - 
 0.09937 i$ &$0.29148 - 0.09895 i$& $0.29456 - 
 0.10768 i$ &$0.32944 - 0.09628 i$
\\
2, 0 &$0.48119 - 0.09741 i$& $0.48328 - 0.09699 i$& $0.48328 - 
 0.09700 i$& $0.48322 - 0.09701 i$& $0.48395 - 0.10057 i$& $0.50632 - 
 0.09612 i$ 
\\
2, 1 &$0.45450 - 0.29924 i$& $0.46001 - 0.29566 i$& $0.46161 - 
 0.29816 i$& $0.46369 - 0.29682 i$& $0.47186 - 0.30943 i$& $0.56110 - 
 0.26022 i$
\\
3, 0 & $0.67489 - 0.09662 i$& $0.67521 - 0.09657 i$& $0.67521 - 
 0.09660 i$& $0.67519 - 0.09660 i$& $0.67546 - 
 0.09846 i$& $0.69173 - 0.09615 i$
 \\
3, 1& $0.65835 - 0.29288 i$& $0.65936 - 0.29243 i$& $0.65969 - 
 0.29320 i$& $0.66048 - 0.29285 i$& $0.66392 - 0.30052 i$& $0.73662 - 
 0.27086 i$
\\
3, 2 &$0.62898 - 0.49442 i$& $0.62721 - 0.49581 i$& $0.63020 - 
 0.49960 i$& $0.63546 - 0.49546 i$& $0.64884 - 0.51250 i$& $0.80100 - 
 0.41515 i$
\\[1ex] 
 \hline
 \end{tabular}
 \caption{QNMs at different orders of the WKB approximation}
\label{table3}
\end{table}

\section{The inverted potential method}
\begin{figure}
\includegraphics[scale=0.45]{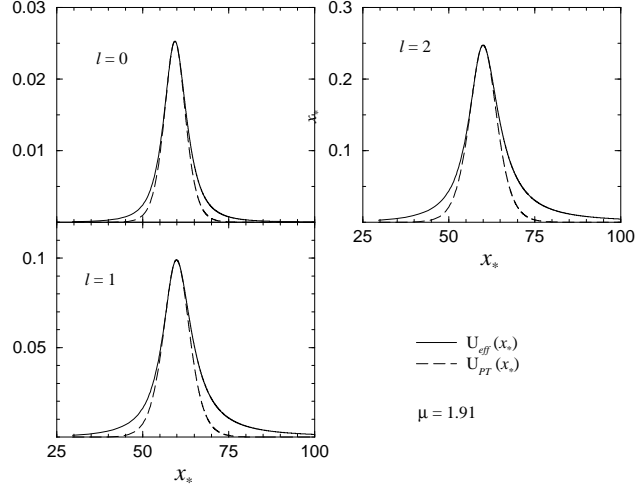}
\caption{Comparison of the P\"{o}schl-Teller and actual effective potential 
for $l$ = 0, 1, 2 and $\mu$ = 1.91.} \label{figPT}
\end{figure}

It has been shown in \cite{INVPOT} that the QNMs of a black hole can be related to the bound states of the so-called inverted black hole potentials. Relying on this property, we approximate the effective potential expressed in the tortoise coordinate by means of the P\"{o}schl-Teller potential \cite{PT} $-U_{PT}(x_{*})$ defined through
\begin{equation}
U_{PT}(x_{*})=\frac{U_0}{\cosh^2{\kappa(x_{*}-x_{*,m})}}
\end{equation}
where $U_0$ and $-2U_0\kappa^2$ denote the height of the maximum and the curvature of the effective potential at the maximum, more precisely
\begin{equation}
\kappa^2=-\frac{1}{2U_0}\left.\frac{d^2U_{eff}}{dx_{*}^2}\right|_{x_{*}=x_{*,m}}=-\frac{K(x_m)}{2U_0},\quad
K(x_m)=f(x_m)\frac{d}{dx}\left.\left(f(x)\frac{d U_{eff}}{dx}\right)\right|_{x=x_{m}}
\end{equation}
with $f(x)$ given by (\ref{f}). Taking into account that the bound states of the P\"{o}schl-Teller potential are given by \cite{INVPOT}
\begin{equation}\label{bs}
\Omega=\kappa\left[-\left(n+\frac{1}{2}\right)+\sqrt{\frac{1}{4}+\frac{U_0}{\kappa^2}}\right],\quad n=0,1,\cdots,N-1
\end{equation}
the corresponding QNMs associated to the black hole potential barrier are obtained by applying the transformation $(U_0,\kappa)\to(U_0,i\alpha)$ to (\ref{bs}). In particular, we have
\begin{equation}\label{QNM_IP}
\sigma_{n\ell}=\sqrt{U_0-\frac{\kappa^2}{U_0}}+i\kappa\left(n+\frac{1}{2}\right),\quad n=0,1,\cdots
\end{equation}
where the dependence on $\ell$ enters through the terms $U_0$ and $\kappa$. Fig. 
\ref{figPT} shows how this P\"{o}schl-Teller (PT) potential compares with the actual 
effective potential for different values of $l$ and $\mu = 1.91$. The PT potential 
reproduces the effective potential quite well in the peak region in all 3 cases. 

In Table~\ref{table:3}, we present some of the QNMs of a noncommutative geometry inspired Schwarzschild black hole computed according to the method described above. According to the method used, the imaginary part of the QNMs do not exhibit any sign flip. Furthermore, it is gratifying to observe that for large values of the parameter $\mu$, the numerical values of the QNMs in the tenth column of Table~\ref{table:3} coincide with the quasinormal frequencies computed by \cite{INVPOT} for the case of the scalar perturbations of the classic Schwarzschild black hole.
\begin{table}[ht]
\centering
 \begin{tabular}{||c| c| c| c| c| c| c|c||} 
 \hline
 $\ell$ & $n$ & $\sigma_{L}$ & $\sigma_{WKB}$ & $\sigma_{e,PT}$ & $\sigma_{PT}$, $\mu=1.95$ & $\sigma_{PT}$, $\mu=2.25$ &$\sigma_{PT}$, $\mu=100$\\ [0.5ex] 
 \hline\hline
 $0$ & $0$ & $0.1105-0.1049i$ & $0.1046-0.1152i$ & $0.1003-0.1232i$ & $0.1022-0.1225i$ & $0.1103-0.1184i$& $0.1148-0.1148i$\\ 
     & $1$ & $0.0861-0.3481i$ & $0.0892-0.3550i$ & $0.1003-0.3697i$ & $0.1022-0.3675i$ & $0.1103-0.3552i$& $0.1148-0.3444i$\\
 $1$ & $0$ & $0.2929-0.0977i$ & $0.2911-0.0980i$ & $0.2983-0.0997i$ & $0.2983-0.1002i$ & $0.2983-0.1009i$& $0.2985-0.1006i$\\
     & $1$ & $0.2645-0.3063i$ & $0.2622-0.3704i$ & $0.2983-0.2991i$ & $0.2983-0.3007i$ & $0.2983-0.3028i$& $0.2985-0.3019i$\\
     & $2$ & $0.2295-0.5401i$ & $0.2235-0.5268i$ & $0.2983-0.4985i$ & $0.2983-0.5012i$ & $0.2983-0.5048i$& $0.2985-0.5032i$\\ 
     & $3$ & $0.2033-0.7883i$ & $0.1737-0.7486i$ & $0.2983-0.6979i$ & $0.2983-0.7017i$ & $0.2983-0.7067i$& $0.2985-0.7045i$\\
 $2$ & $0$ & $0.4836-0.0968i$ & $0.4832-0.0968i$ & $0.4881-0.0949i$ & $0.4878-0.0957i$ & $0.4873-0.0977i$& $0.4873-0.0979i$\\
     & $1$ & $0.4639-0.2956i$ & $0.4632-0.2958i$ & $0.4881-0.2847i$ & $0.4878-0.2873i$ & $0.4873-0.2932i$& $0.4873-0.2937i$\\
     & $2$ & $0.4305-0.5086i$ & $0.4317-0.5034i$ & $0.4881-0.4745i$ & $0.4878-0.4788i$ & $0.4873-0.4887i$& $0.4873-0.4895i$\\
     & $3$ & $0.3939-0.7381i$ & $0.3926-0.7159i$ & $0.4881-0.6643i$ & $0.4878-0.6704i$ & $0.4873-0.6842i$& $0.4873-0.6853i$
     \\[1ex] 
 \hline
 \end{tabular}
 \caption{Normal modes for scalar perturbations of the noncommutative geometry inspired Schwarzschild metric. The third and fourth columns represent  the numerical values found by \cite{Leaver} and \cite{IR}, respectively, for a classic Schwarzschild black hole. Here, $\sigma_{PT}$ is the dimensionless frequency computed according to (\ref{QNM_IP}), and $\sigma_{e,PT}$ denotes the QNM in the case of an extreme non-commutative geometry inspired Schwarzschild black hole.}
\label{table:3}
\end{table}

\section{QNMs by the Asymptotic Iteration Method (AIM)}
From the above exposition it becomes evident that it is important to choose an 
appropriate method to determine the QNMs. It is worthwhile to look for other similar 
cases where conventionally used methods fail. 

In this context, it is interesting to observe that the new branch of QNMs for a 
massless scalar field in the classic Schwarzschild metric derived analytically 
in \cite{DavideB1,Panosso1,DavideB2} can also be obtained numerically by using the 
improved AIM \cite{AIM}. In order to apply this method, we
must consider an ad hoc initial ansatz for a massless scalar field by incorporating its behaviour at the horizon and at infinity as we did in equation (12) in \cite{DavideB1}. This gives rise to the homogeneous linear second order differential equation
\begin{equation}\label{DGL}
f^{''}_{\sigma\ell}=\lambda_0(x)f^{'}_{\sigma\ell}+s_0(x)f_{\sigma\ell}
\end{equation}
where
\begin{equation}
\lambda_0(x)=c-1+\frac{2c-1}{x}-\frac{c}{x-1},\quad
s_0(x)=-\frac{d}{x}+\frac{d}{x-1}-\frac{c^2}{x^2},\quad c=1-4i\sigma,\quad d=c^2+(c-1)^2+\ell(\ell+1).
\end{equation}
The implementation of the improved AIM requires first to differentiate (\ref{DGL}) $n$ times with respect to the independent variable, which produces the following equation
\begin{equation}
f^{(n+2)}_{\sigma\ell}=\lambda_n(x)f^{'}_{\sigma\ell}+s_n(x)f_{\sigma\ell}
\end{equation}
with
\begin{eqnarray}
\lambda_n(x)&=&\lambda^{'}_{n-1}(x)+s_{n-1}(x)+\lambda_0(x)\lambda_{n-1}(x),\label{j1}\\
s_n(x)&=&s^{'}_{n-1}(x)+s_{0}(x)\lambda_{n-1}(x).\label{j2}
\end{eqnarray}
Then, $\lambda_n$ and $s_n$ undergo a Taylor series expansion 
\begin{equation}
\lambda_n(x)=\sum_{i=0}^\infty c_n^i(x-x_0)^i,\quad s_n(x)=\sum_{i=0}^\infty d_n^i(x-x_0)^i
\end{equation}
around some point, $x_0$. By replacing the above expansions in (\ref{j1}) and (\ref{j2}) a set of recursion relations can be derived (see details in \cite{AIM}) from which one obtain the quantization condition
\begin{equation}\label{QCAIM}
d^0_n c^0_{n-1}-d^0_{n-1}c^0_n=0.
\end{equation}
We solved (\ref{QCAIM}) to find the QNMs. In Table~\ref{table:x}, we show the fundamental quasi-normal frequencies for a massless scalar field in the classic Schwarzschild metric with $\ell=0,1,2$. It is worth mentioning that a number of $25$ iterations was employed for the improved AIM method. Last but not least, the same method was adopted in \cite{Cho} to obtain the QNMs computed by \cite{Leaver} with the continued fraction method.
\begin{table}[ht]
\centering
 \begin{tabular}{||c| c| c||} 
 \hline
 $\ell$ & $\sigma$, $n=0$ & $\mbox{Exact}$\\ [0.5ex] 
 \hline\hline
 $0$ & $0.1250-0.1250i$ & $0.1250-0.1250i$ \\ 
 $1$ & $0.2795-0.1250i$ & $0.2795-0.1250i$ \\
 $2$ & $0.4506-0.1250i$ & $0.4506-0.1250i$ 
     \\[1ex] 
 \hline
 \end{tabular}
 \caption{Fundamental normal modes for massless scalar perturbations of the classic Schwarzschild metric. The second column reports the fundamental quasi-normal frequencies obtained by means of the improved AIM while the third column displays the exact fundamental QNMs computed using formula (52) in \cite{DavideB1}.}
\label{table:x}
\end{table}

\section{Conclusions}
Instabilities of microscopic black holes are an interesting area of research as certain 
examples in the literature show. The existence of such instabilities could possibly 
hint toward the inappropriateness of models of microscopic black holes. 
We choose here, as an example, the case of a Schwarzschild black hole inspired by 
noncommutative geometry. Indeed one of our motivation to probe into this case 
is a hint of instability revealed at the third order of the WKB method. 
It is then obligatory to pursue further the nature of such a possibility, either 
by assuring the convergence of the WKB method or, in case the convergence fails, 
choosing another appropriate algorithm to determine the QNMs. We show by going 
up to the sixth order of WKB, that the convergence of the WKB method is not guaranteed 
for the cases under discussion. The inverted potential method is then a more suitable 
alternative which reveals that these instabilities were only an artifact of the 
third order WKB (see \cite{hatsuda} for a discussion on the 
divergence of the WKB series). Furthermore, we show how the asymptotic iterative method confirms the existence of a 
new branch of analytical QNMs recently discovered in \cite{DavideB1}.

\appendix
\section{Derivation of the inequality (\ref{ungl})}\label{ineq}
We start by observing that the integral in (\ref{inizio}) is positive and bounded. Switching back to the coordinate $x$ and using (\ref{def_tor}) together with the second relation in (\ref{tor_eq}), the integral (\ref{inizio}) becomes
\begin{eqnarray}
\mathcal{I}&=&\int^{+\infty}_{-\infty}\mathcal{U}_{eff}(x_{*})dx_{*}=\int^{+\infty}_{x_h}\left[\frac{1}{x}\frac{df}{dx}\frac{dx}{dx_{*}}+\frac{\lambda}{x^2}f(x)\right]\frac{dx_{*}}{dx}dx=\int^{+\infty}_{x_h}\left[\frac{1}{x}\frac{df}{dx}f(x)+\frac{\lambda}{x^2}f(x)\right]\frac{dx}{f(x)},\\
&=&\int^{+\infty}_{x_h}\left[\frac{1}{x}\frac{df}{dx}+\frac{\lambda}{x^2}\right]dx=\frac{\lambda}{x_h}+\mathcal{J},\quad\mathcal{J}=\int^{+\infty}_{x_h}\frac{1}{x}\frac{df}{dx}~dx,\quad\lambda=\ell(\ell+1).\label{pip}
\end{eqnarray}
Furthermore, (\ref{f}) and relation 7.1.1 in \cite{abra} give
\begin{equation}
\frac{df}{dx}=\frac{2}{x^2}\mbox{erf}\left(\frac{1}{2}\mu x\right)-\frac{2\mu}{\sqrt{\pi}x}e^{-\frac{1}{4}\mu^2 x^2}-\frac{\mu^3}{\sqrt{\pi}}xe^{-\frac{1}{4}\mu^2 x^2},
\end{equation}
and the integral $\mathcal{J}$ in (\ref{pip}) can be rewritten as
\begin{equation}\label{Ical}
\mathcal{J}=2\int^{+\infty}_{x_h}\frac{1}{x^3}\mbox{erf}\left(\frac{1}{2}\mu x\right)dx-\frac{2\mu}{\sqrt{\pi}}\int^{+\infty}_{x_h}\frac{1}{x^2}e^{-\frac{1}{4}\mu^2 x^2}dx-\frac{\mu^3}{\sqrt{\pi}}\int^{+\infty}_{x_h}e^{-\frac{1}{4}\mu^2 x^2}dx.
\end{equation}
The last integral in the above expression can be immediately computed using the definition of the error function together with the property $\mbox{erf}(s)\to 1$ as $s\to+\infty$. More precisely, we find
\begin{equation}\label{A1}
\int^{+\infty}_{x_h}e^{-\frac{1}{4}\mu^2 x^2}dx=\frac{\sqrt{\pi}}{\mu}-\frac{\sqrt{\pi}}{\mu}\mbox{erf}\left(\frac{1}{2}\mu x_h\right).
\end{equation}
The second integral in (\ref{Ical}) can be easily computed integrating by parts and using again the definition of the error function. This leads to the result
\begin{equation}\label{A2}
\int^{+\infty}_{x_h}\frac{1}{x^2}e^{-\frac{1}{4}\mu^2 x^2}dx=-\frac{\mu\sqrt{\pi}}{2}+\frac{e^{-\frac{1}{4}\mu^2 x_h^2}}{x_h}+\frac{\mu\sqrt{\pi}}{2}\mbox{erf}\left(\frac{1}{2}\mu x_h\right).
\end{equation}
Finally, the first integral in (\ref{Ical}) can be evaluated using 1.5.1.8 in \cite{Prud} with $n=1$ and $a=\mu/2$ there. We obtain
\begin{equation}\label{A3}
\int^{+\infty}_{x_h}\frac{1}{x^3}\mbox{erf}\left(\frac{1}{2}\mu x\right)dx=-\frac{\mu^2}{4}+\frac{\mu}{2\sqrt{\pi}x_h}e^{-\frac{1}{4}\mu^2 x_h^2}+\frac{2+\mu^2 x_h^2}{4x^2_h}\mbox{erf}\left(\frac{1}{2}\mu x_h\right).
\end{equation}
Replacing (\ref{A1}), (\ref{A2}), and (\ref{A3}) into (\ref{Ical}) gives
\begin{equation}\label{jota}
\mathcal{J}=-\frac{\mu^2}{2}-\frac{\mu}{\sqrt{\pi}x_h}e^{-\frac{1}{4}\mu^2 x_h^2}+\frac{2+\mu^2 x_h^2}{2x^2_h}\mbox{erf}\left(\frac{1}{2}\mu x_h\right).
\end{equation}
If we substitute (\ref{jota}) into (\ref{pip}) and recall that the integral $\mathcal{I}$ is positive, we obtain the inequality 
\begin{equation}
\frac{\lambda}{x_h}-\frac{\mu^2}{2}-\frac{\mu}{\sqrt{\pi}x_h}e^{-\frac{1}{4}\mu^2 x_h^2}+\frac{2+\mu^2 x_h^2}{2x^2_h}\mbox{erf}\left(\frac{1}{2}\mu x_h\right)>0
\end{equation}
which can be cast into the form of (\ref{ungl}).

\acknowledgments
The authors would like to thank the anonymous referee for his/her enlightening 
comments on the manuscript and his/her valuable suggestions. The authors are grateful 
to R. A. Konoplya for providing the code to evaluate the QNMs using the WKB method. 
One of the authors (N.G.K.) thanks the Faculty of Science, Universidad de los
Andes, Colombia for financial support through grant no. P18.160322.001-17.

\end{document}